\documentclass[12pt]{article}
 \usepackage[dvips]{graphicx}
 \usepackage{amsmath}
 \usepackage{amsxtra}
 \usepackage{amssymb}
\begin{document}
\title {Transition From Quantum To  Classical Mechanics As Information Localization}
\author{ A.Granik\thanks{Department of Physics, University of the
Pacific,Stockton,CA.95211} } \maketitle
\begin{abstract}
Quantum parallelism implies a spread of information over the space
in contradistinction to  the classical mechanical situation where
the information is "centered" on a fixed trajectory of a classical
particle. This means that a quantum state becomes specified by
more indefinite data. The above spread resembles, without being an
exact analogy, a transfer of energy to smaller and smaller scales
observed in the hydrodynamical turbulence. There, in spite of the
presence of dissipation (in a form of kinematic viscosity), energy
is still conserved. The analogy with the information spread in
classical to quantum transition means that in this process the
information is also conserved. To illustrate that, we show (using
as an example a specific case of a coherent quantum oscillator)
how the Shannon information density continuously changes in the
above transition . In a more general scheme of things, such an
analogy allows us to introduce a "dissipative" term (connected
with the information spread) in the Hamilton-Jacobi equation and
arrive in an elementary fashion at the equations of classical
quantum mechanics (ranging from the Schr\"{o}dinger to
Klein-Gordon equations). We also show that the principle of least
action in quantum mechanics is actually the requirement for the
energy to be bounded from below.
\end{abstract}
Keywords: {\small Classical to quantum transition; information density transformation}\\

\section{INTRODUCTION}

Present day efforts in making quantum computing a reality are
centered mainly on harnessing immense parallelism (e.g., entangled
states) inherent in quantum mechanics. In terms of information
content such a parallelism means that information is spread over
the whole space. The implication is that information density is
not delta-function-like ( as in a classical case) but is
represented by a  'broader' function. In a sense, this can be
interpreted as an  information spread, in contradistinction to a
classical case, where the information is centered around the
well-defined path determined by the classical equations of motion.
The problem of extracting the information so spread becomes
central to every possible quantum computer. Therefore it seems
important
to determine how  this spread of information occurs in general.\\

For the first time an idea about a "spread" of information in a
transition from the classical to the quantum world was expressed
by P.Dirac more than 70 years ago \cite{PD}. He wrote, "The
limitation in the power of observation puts a limitation on the
number of data that can be assigned to a state. Thus a  state of
{\bf an atomic system} must be specified by fewer or {\bf more
indefinite} data than a complete set of numerical values for all
the coordinates and velocities at some instant of time."
Unfortunately, he did not
elaborate further on this idea.\\

Does all this mean that information is lost via some sort of
dissipation in a transition from classical to quantum case? In
another words, is information lost in a literal sense of the word,
or simply 'spread around', that is the respective information
density undergoes a change in a transition from a classical to
quantum case? The mechanism of the latter represents what we would
call the $information~spread$. As will be shown below, this
$information~spread$ is the correct answer.\\

A tentative approach to find the answer to this problem  in a
general way was outlines at \cite{AG2}. There we observed that
contrary to the conventional point of view (regarding the
transition from classical to quantum physics as being necessarily
due to decoherence \cite{WZ}), our investigation of a superfluid
state demonstrated coherence preservation. Indeed, in our view
decoherence plays essentially no role in the transition from
ordinary classical physics to quantum physics. This transition can
occur in a continuous fashion
preserving the coherence in a classical state.\\

We also argued that "Whereas the entropy of any deterministic
classical system described by a principle of least action is zero,
one can assign a "quantum information" to quantum mechanical
degree of freedom equal to Hausdorff area of the {\bf deviation}
from a classical path." This raises an interesting problem of
realization of a quantum computer based on a continuous transition
from a quantum coherent state to a classical coherent state. Such
an approach is contrary to the conventional treatment of quantum
computing where quantum coherence is destroyed by classical
measurements. The difficulty of preserving quantum coherence
lies at the heart of the general difficulty of realizing such a computer.\\

In what follows we demonstrate (using a coherent state of a
quantum oscillator) how the information-preserving mechanism,
characterized by a spatial spread of information density, occurs.
In a sense (and only in a sense), this mechanism is analogous to
the effect of dissipation on the velocity profile of a viscous
fluid, illustrated, for example, by Stokes's first problem about a
suddenly accelerated plane wall immersed in a viscous fluid \cite
{HS}. We write " in a sense", since in contradistinction to fluid
mechanics ( where the system dissipates energy), here  no loss
of information occurs.\\

Such an analogy allows us to show how in a $general$ scheme of
things (not restricted to some special cases as in \cite{MB}) the
addition of the specific "dissipative" term (similar to the
dissipative term in fluid mechanics) to the classical equations of
motion will lead in a natural way to the wave equations, ranging
from the Schr$\mathrm{\ddot{o}}$dinger to Klein-Gordon, to Dirac
equations \footnote{It is interesting  that for the $2+1$-
dimensional case a certain transformation \cite{RK} reduces the
Schr$\mathrm{\ddot{o}}$dinger equation to a pair of differential
equation, one of which  is the Navier-Stokes vorticity equation}.
\\

If we consider the Shannon information for the coherent state of a
quantum oscillator then we will be able to explicitly illustrate
how the information density associated with this oscillator
continuously changes from a function spread over the  whole
spatial domain in quantum case to the delta-function centered on
the domain occupied by the  values of the spatial coordinate (that
is $x=acos\omega t$) allowed by classical mechanics. In
particular, this proves (in full agreement with the above
arguments) that in fact the $total$ information is not lost, but
rather a change of its space density  occurs.\\

\section{Information Density in Classical and Quantum Regimes}
Let us consider the Shannon information $I$
\begin{equation}
\label{S} I=-\sum_{i=0}^Np_ilog_2p_i
\end{equation}
Here $p_i$ is the probability of an event $A_i$. In what follows
we replace $log_2$ by the natural logarithm which is not going to
change the meaning of the information, but will simply introduce a
non-essential numerical factor. For our purposes we define the
probability in (\ref{S}) with the help of  the probability density
function $\Psi(x,t)$, as it is used in quantum mechanics.\\

In this context the probability $p_i$ (defining a probability of
finding a particle in the space interval $x_i,x_{i+1}$) can be
written  as follows
\begin{equation}
\label{S1} p_i=\int_{x_i}^{x_{i+1}}|\Psi|^2dx
\end{equation}
Therefore
\begin{equation}
\label{S2}p_iLnp_i=\int_{x_i}^{x_{i+1}}|\Psi|^2dx~Ln(\int_{x_i}^{x_{i+1}}|\Psi|^2dx)
\end{equation}
For a coherent state of a quantum oscillator (\ref{S2}) yields:
\begin{equation}
\label{S3}
p_iLnp_i=\\
\frac{\alpha}{\sqrt{\pi}}\int_{x_i}^{x_{i+1}}e^{\alpha^2(\tilde{x}-cos\omega
t)^2}d\tilde{x}Ln\{\alpha\int_{x_i}^{x_{i+1}}e^{\alpha^2(\tilde{x}-cos\omega
t)^2}d\tilde{x}\}
\end{equation}
where $\alpha = a\sqrt{{m\omega}/{\hbar}}$, $m$ is particle's
mass, $a$ is the classical amplitude,  $\omega$  is the classical
frequency and $\tilde{x}=x/a$ is the dimensionless coordinate.
Parameter $\alpha$ has a clear physical meaning. Since
\begin{equation}
\label{add} \frac{\hbar}{ma\omega}=\lambda_{db}
\end{equation}
(where $\lambda_{db}$ is the respective DeBroglie
wavelength),$$\alpha= \frac{a}{\lambda_{db}}$$ indicates whether
particle dynamics is a classical ($\alpha \gg1$) or a quantum one
($\alpha\sim 1$).\\

 Integrating (\ref{S3}), we obtain
\begin{equation}
\label{S4} p_iLnp_i=\frac{1}{2}[\Phi(\alpha y_{i+1})-\Phi(\alpha
y_i)]Ln\{[\Phi(\alpha y_{i+1})-\Phi(\alpha y_i)]\}
\end{equation}
Here $y=\tilde{x}-cos\omega t$ and
$$\Phi(y)=\frac{2}{\sqrt{\pi}}\int_0^ye^{-z^2}dz$$\

We consider a situation where $$y_{i+1}-y_i=\Delta y_i\ll 1$$

Inserting this in (\ref{S4}) we arrive at the following
\begin{equation}
\label{S5}p_iLnp_i=-\frac{1}{\sqrt{\pi}}e^{-(\alpha
y_i)^2}\{1+\frac{Ln\pi}{2}+(\alpha y_i)^2\}\alpha\Delta
y_i+O[(\Delta y_i)^2]
\end{equation}
where we use $$xLnx_{(x\rightarrow 0)}\rightarrow-x$$ Therefore
(\ref{S}) becomes
\begin{equation}
\label{S6} -\sum_{i=0}^N
p_iLnp_i=\frac{1}{\sqrt{\pi}}\sum_{i=0}^Ne^{-(\alpha
y_i)^2}\{1+\frac{Ln\pi}{2}+(\alpha y_i)^2\}\alpha\Delta y_i
\end{equation}
 In the limit $N\rightarrow\infty,\Delta
y_i\rightarrow dy$ relation (\ref{S6}) yields the following
integral
\begin{equation}
\label{S7} -\sum_{i=0}^N
p_iLnp_i=\frac{1}{2\sqrt{\pi}}\int_{-\infty}^{\infty}e^{-(\alpha
y_i)^2}\{1+\frac{Ln\pi}{2}+(\alpha y_i)^2\}\alpha dy
\end{equation}
Therefore the integral function
\begin{equation}
\label{S8}\frac{dI}{dx}=\frac{\alpha}{2\sqrt{\pi}}  e^{-(\alpha
y_i)^2}\{1+\frac{Ln\pi}{2}+(\alpha y_i)^2\}
\end{equation}
represents the $space ~information ~density$, that is the
information (per unit of dimensionless length) about finding the
particle at a certain location $x$. The graph of this function for
various values of $a/\lambda_{db}=1,2,...,10$ is shown in Fig.$1$.
\begin{figure}
 \begin{center}
\includegraphics[width=6cm, height=6cm]{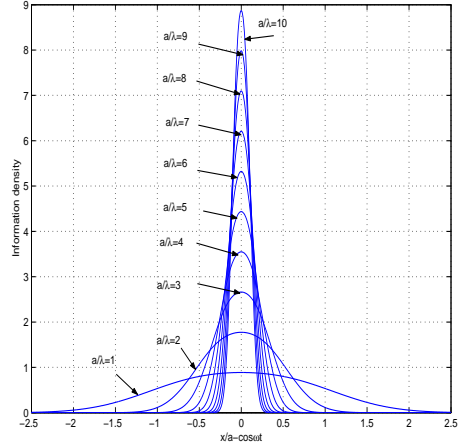}
\caption{\small  Information density ( per unit of dimensionless
length $\tilde{x}$) as a function of $\tilde{x}-cos\omega t$ for
different values of $a/\lambda_{db} =1,2,..., 10$}
 \end{center}
 \end{figure}\\

In the limit of $\alpha=a/\lambda_{db}\rightarrow\infty$ the space
information density  tends to the delta-function. This indicates
the onset of a purely classical regime, such that outside the
region $x=acos\omega t$ (occupied by the displacement of the
classical oscillator) the information density is 0. Thus all the
information is "concentrated" in the region $x-acos\omega t=0$.\\
\begin{figure}
 \begin{center}
\includegraphics[width=6cm, height=6cm]{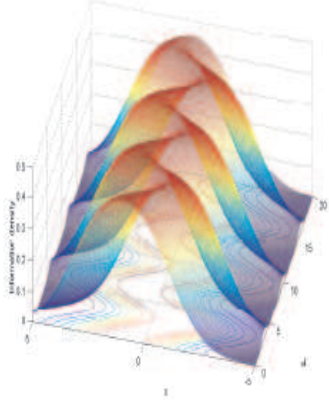}
\caption{\small  Information density ( per unit of dimensionless
length $\tilde{x}$) as a function of $\tilde{x}$ and $t$ for
$a/\lambda_{db} =.5$}
 \end{center}
 \end{figure}

In the opposite limit $\alpha=a/\lambda_{db}\rightarrow 1$ the
information density function is spread over  the domain
$-\infty<y<\infty$ of all possible values of $x-acos\omega t$
reaching outside the region occupied by the displacement of the
classical oscillator. This indicates a quantum regime
characterized by the information  which is not "concentrated" on a
well defined path (of measure zero) but is rather "diffused".
This, of course, does not mean that the information  is lost. On
the contrary, the
information is preserved, being however "spread" over the whole space.\\

It is instructive to provide the graphs of the information density
as a function of the spatial and temporal coordinates. These
graphs are presented in Figures $2$, $3$, and $4$ for the ratios
$a/\lambda_{bd}=0.5,5,15$ respectively.
\begin{figure}
 \begin{center}
\includegraphics[width=6cm, height=6cm]{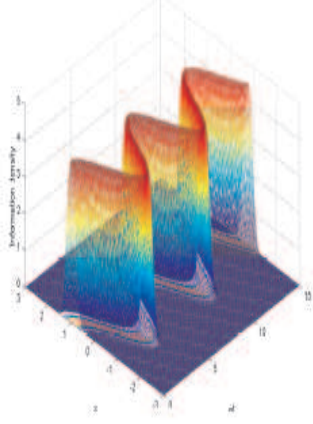}
\caption{\small  Information density ( per unit of dimensionless
length $\tilde{x}$) as a function of $\tilde{x}$ and $t$ for
$a/\lambda_{db} =5$}
 \end{center}
 \end{figure}
 Once again, one can easily see that with the increase of the ratio $a/\lambda_{db}$,
 that is the approach to the classical regime, the information
 density tends to be concentrated along the classical path
 $\tilde{x}=cos(\omega t)$.\\

Here we must emphasize that the spatial information spread
expressed in terms of $\bf spatial$ information density refers
$exclusively$ to the Schr$\mathrm{\ddot{o}}$dinger representation
of quantum mechanics. It describes the respective dynamics in
spatial-temporal terms with the help of the quantum "potential",
the wave function $\Psi(x,t)$.  On the other hand, the equivalent
second quantization representation of quantum mechanics deals only
with the number states, without reference to their spatial
distribution. Therefore it is important to find out how the
respective information density varies with changes in number
states, which can be quite different from the changes of
information density in the Schr$\mathrm{\ddot{o}}$dinger representation.\\

\subsection{Information Density for a Coherent State of the Quantum Oscillator in the Number State
Representation}

Let us calculate this information density. The probability to find
an oscillator in the $n-th$ state is
\begin{equation}
\label{n1} P(n,<n>)=\frac{<n>^n}{n!}e^{-<n>}
\end{equation}
where the average number of states
\begin{equation}
\label{n11}
 <n>=\frac{(m\omega^2<x>^2+<p^2>/m)/2}{\hbar\omega}
 \end{equation}
The Shannon information is then
\begin{equation}
\label{n2}
I=-\sum_nP_nLnP_n=<n>(1-Ln<n>)+e^{-<n>}\sum_n\frac{<n>^n}{n!}Ln(n!)
\end{equation}
where we use natural  logarithm, instead the one base $2$, which
would introduce into the result a nonessential numerical factor.
From Eq.(\ref{n2}) follows
\begin{equation}
\label{n3}
\frac{dI}{d<n>}=e^{-<n>}\sum_n\frac{<n>^n}{n!}Ln(n+1)-Ln(<n>)
\end{equation}
In general, the sum in (\ref{n3}) cannot be found in a closed
form. However, we can evaluate it in the quantum limit $<n>\ll 1$
\begin{equation}
\label{n4} \lim_{n\ll 1}=-Ln<n>\gg 1
\end{equation}
Since there is no analytical solution to (\ref{n3}), the numerical
evaluation allows us to represent the result as a graph
$dI/d<n>=f(<n>)$. It is shown in  Fig. $5$. One can easily see
that the {\it number state} information density $decreases$ in a
transition from a quantum to a classical regime, in
contradistinction to the $spatial$ information density with its
sharp increase around the classical trajectory.\\

The apparent paradox is resolved by observing that in the number
state representation the Shannon information is not conserved
anymore. In fact, according to (\ref{n2}) it increases in a
transition from the quantum to the classical case, since the
average number of states given by \ref{n11} (i.e. roughly the
ratio of a classical amplitude and the respective De Broglie
wavelength) monotonically increases. Therefore the two
representations conceptually differ in this respect.
\begin{figure}
 \begin{center}
\includegraphics[width=6cm, height=6cm]{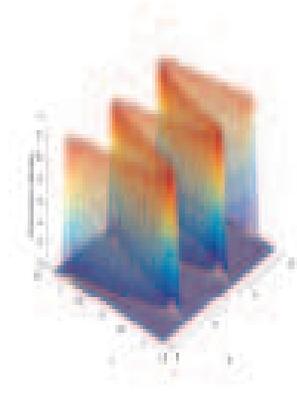}
\caption{\small  Information density ( per unit of dimensionless
length $\tilde{x}$) as a function of $\tilde{x}$ and $t$ for
$a/\lambda_{db} =15$}
 \end{center}
 \end{figure}\
\begin{figure}
 \begin{center}
\includegraphics[width=6cm, height=6cm]{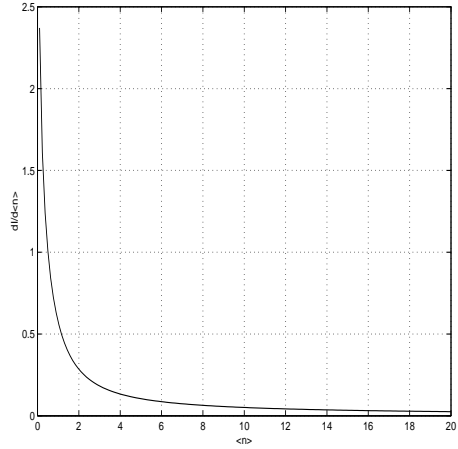}
\caption{\small Number state information density as a function of
the average number of states $<n>$}
 \end{center}
 \end{figure}\

\section{Schr$\mathrm{\ddot{o}}$dinger Equation as a Result of Information Spread}
The previous sections imply that a judicious introduction of a
"dissipative" (or rather quasi-dissipative) term (signaling a
$spatial$ spread of information) into the equations of classical
mechanics can result in the respective quantum equations. To
achieve this goal we use the following experimental facts:\

1) Quantum phenomena are characterized by the superposition
principle, implying that in contradistinction to the classical
mechanics with its non-linear equations, the respective quantum
equations must be linear.\

2)There exists a smallest finite quantum of energy
$E=\hbar\omega$, which in the phase space corresponds to the
finite elemental area $\hbar$.

3) Quantum phenomena exhibit both particle and wave properties.\\

We begin with the second law of Newton for a single particle
moving from p.$A$ to p.$F$ (see Fig.$6$). The particle can do that
by taking any possible path connecting these two points. Therefore
for any fixed moment of time, say $t=1$ particle's momentum would
depend on the spatial coordinate, that is
$\vec{p}=\vec{p}(\vec{x},t)$. This means that now the substantial
derivative $d/dt=\partial/\partial t +v_j\partial/\partial x_j$.
In a sense, instead of watching the particle evolution in time one
watches the evolution of its momentum  in space and time. This
situation is analogous to the Euler's description of motion of a
fluid (an alternative to the Lagrange description). The other way
to look at that is to consider a "flow" of an "elemental" path and
describe its "motion" in terms of  its coordinates and velocity.
\begin{figure}
 \begin{center}
 \includegraphics[width=6cm, height=6cm]{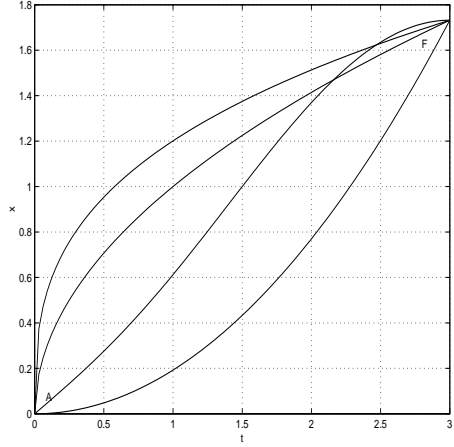}
 \caption{\small A few paths of a path set connecting the initial and the final points
traveled by a particle in $t= 3sec$. It is clearly seen that
particle's  velocity ( momentum) is a function of both coordinate
$x$ and time $t$}
 \end{center}
 \end{figure}
Taking this into account we write the second law as follows
(e.g.,\cite{AG})
\begin{equation}
\label{30} \frac{d p_j}{dt}=-\frac{\partial\Pi_{jk}}{\partial x_k}
\end{equation}
where $\Pi_{jk}$ is the momentum flux density tensor, and we adopt
the convention of summation over the repeated indices.

In a purely classical case $$\Pi_{jk}=U\delta_{jk}$$ where $U$ is
the potential, representing an absence of  "friction" between
different possible paths. On the other hand, at the micro-level we
postulate a "viscous" transfer of momentum from a path with a
greater momentum to  paths with a smaller momentum, similar to a
transfer of energy from larger to smaller scales in a turbulent
motion. Therefore we add to the "ideal" momentum flux $P_{jk}$ in
(\ref{30}) a term analogous to the one used in classical mechanics
of  fluids. This yields the following expression for
$\Pi_{jk}$(e.g.,\cite{LL})
\begin{equation}
\label{31} \Pi_{jk}=U\delta_{jk}- \nu_1^q(\frac{\partial
p_j}{\partial x_k}+\frac{\partial p_k}{\partial
x_j}-\frac{2}{3}\delta_{jk}\frac{\partial p_l}{\partial
x_l})-\nu_2^q\delta_{jk}\frac{\partial p_l}{\partial x_l}
\end{equation}
where "viscosities" $\nu_1^q,\nu_2^q$ will be determined in what
follows.\\

Inserting (\ref{31}) in (\ref{30}) we obtain in vector notations:
\begin{equation}
\label{32} \frac{\partial\vec{p}}{\partial
t}+\frac{1}{m}(\vec{p}\centerdot\nabla)\vec{p}=-\nabla
U+\nu_1^q\nabla^2\vec{p}+(\frac{1}{3}\nu_1^q+\nu_2)\nabla
div\vec{p}
\end{equation}

Application of $curl$ to both sides of (\ref{32}) results in the
following:
\begin{equation}
\label{33} \frac{\partial}{\partial
t}\nabla\times\vec{p}-\frac{1}{m}\nabla\times[\vec{p}\times(\nabla\times\vec{p})]
-\nu_1^q\nabla\times[\nabla\times(\nabla\times\vec{p})]=0
\end{equation}
Equation (\ref{33}) is identically satisfied if
$\nabla\times\vec{p}=0$, or equivalently
\begin{equation}
\label{34} \vec{p}=\nabla S^{(q)}
\end{equation}
where $S^{(q)}$ is a new "effective action".\\

It must be said, that this "effective action" serves only as an
interim auxiliary function without a clear physical meaning,
which  allows us to make a transition to the quantum case.
Importantly enough, in contradistinction to the conventional
hydrodynamical treatment of viscous fluid, the present case is
irrotational. In conventional hydrodynamics of incompressible
viscous fluids (with $div\vec{v}=0$) motion with $curl\vec{v}=0$
represents a potential motion $\nabla^2\vec{v}=0$. However at the
atomic scales $\nabla^2 S\neq 0$, because of the absence of
continuity equation analogous to the one in incompressible fluid,
that is now $div\vec{p}\neq 0$\\

Substituting(\ref{34}) in (\ref{32}) and using the vector
identities
$$(\vec{u}\centerdot\nabla)\vec{u}\equiv\frac{1}{2}[\nabla
(\vec{u}\centerdot\vec{u})-\vec{u}\times curl\vec{u}]$$
$$\nabla^2\vec{u}\equiv\nabla(\nabla\centerdot\vec{u})-\nabla\times(\nabla\times\vec{u})$$

we obtain the following equation
\begin{equation}
\label{35} \nabla\{\frac{\partial S^{(q)}}{\partial
t}+\frac{1}{2m}(\nabla S^{(q)})^2+U-\nu^q\nabla^2 S^{(q)}\}=0
\end{equation}
where $$\nu^q=\frac{4}{3}\nu_1^q+\nu_2^q$$ Equation (\ref{35}) is
identically satisfied if
\begin{equation}
\label{35a} \frac{\partial S^{(q)}}{\partial
t}+\frac{1}{2m}(\nabla S^{(q)})^2+U=\nu^q\nabla^2 S^{(q)}
\end{equation}
We have arrived at what can be called a modified Hamilton-Jacobi
equation with "dissipation". As we have already indicated, it does
not play any role at the macro-scales of classical mechanics due
to the smallness of the dissipative term as compared to the rest
of the terms. As will be shown later, this smallness is directly
related to the ratio of the DeBroglie wavelength and the
characteristic length on a classical scale. \\

Since the obtained equation is non-linear, it cannot be used to
describe quantum phenomena, since this contradicts the
experimental facts about superposition of quantum states. In
addition, (\ref{35a}) does not have a wave solution, which again
contradicts the experimental facts about quantum phenomena.
Therefore we have (if possible) to identically transform
(\ref{35a}) into an equation which would be\
\begin{itemize}
 \item a) linear\

 and\

\item b) would allow wave solutions.
\end{itemize}

 Requirement $b)$ can be achieved (at least for a time dependence),
 if it would be possible to transform (\ref{35a}) into
 a homogeneous (but still nonlinear) partial differential equation  of order
 $2$. To test this proposition we introduce a new function, say
 $\Psi(\vec{x},t)$, such that
 \begin{equation}
 \label{S10}
 S^{(q)}=S^{(q)}(\Psi)
 \end{equation}
 Inserting (\ref{S10}) in (\ref{35a}) we obtain:
 \begin{equation}
 \label{S11}
 \frac{dS^{(q)}}{d\Psi}\frac{\partial\Psi}{\partial
 t}+\frac{1}{2m}(\frac{dS^{(q)}}{d\Psi})^2(\nabla\Psi)^2+U=\nu^{(q)}
 [\frac{dS^{(q)}}{d\Psi}(\nabla\Psi)^2+\frac{d^2S^{(q)}}{d\Psi^2}\nabla^2\Psi]
 \end{equation}\

 Amazingly enough, this equation becomes a homogenous nonlinear partial differential equation
 of order $2$ with respect to the new function $\Psi$, if and only if
 the functional dependence (\ref{S10}) is as follows:
 \begin{equation}
 \label{S12}
 \frac{dS^{(q)}}{d\Psi}=\frac{A}{\Psi}
 \end{equation}\

Solving (\ref{S12}) we obtain:
\begin{equation}
\label{S13} S^{(q)}= ALn\Psi+B
\end{equation}
where constant $A$ a will be determined later with the help of the
requirements formulated at the beginning of this section. Since
constant $B$ does not enter into the resulting equation with
respect to function $\Psi$, we set it equal to 0 without any loss
of generality. Therefore (\ref{S13}) yields
\begin{equation}
\label{S13a}
 S^{(q)}=ALn\Psi
\end{equation}
This relation is exactly what Schr$\mathrm{\ddot{o}}$dinger
originally introduced "by hand"  in his first paper in the
historical series of $6$
papers on the wave equation \cite{S1}.\\

Meanwhile we substitute (\ref{S12}) in (\ref{S11}) and obtain:
\begin{equation}
\label{S14} \frac{1}{A}\Psi\frac{\partial\Psi}{\partial
t}+\frac{1}{2m}(\nabla\Psi)^2+\frac{1}{A^2}U\Psi^2=\frac{1}{A}\nu^{(q)}[\Psi\nabla^2\Psi-
(\nabla\Psi)^2]
\end{equation}
It is clear that for a particular case of the function $U$ being
time-independent, (\ref{S14}) allows a solution proportional to
$exp(i\omega t)$.\\

To convert (\ref{S14}) into a linear equation we have to "get rid"
of the nonlinear term $(\nabla\Psi)^2$. Since the "viscosity"
$\nu^{(q)}$ was introduced in such a way that its exact value was
undetermined, we can use this fact and eliminate the nonlinear
term by the appropriate choice of $\nu^{(q)}$. This procedure
yields:
\begin{equation}
\label{S15} \nu^{(q)}=\frac{A}{2m}
\end{equation}
As a result, equation (\ref{S14}) becomes
\begin{equation}
\label{S16} A\frac{\partial\Psi}{\partial
t}-\frac{A^2}{2m}\nabla^2\Psi+U\Psi=0
\end{equation}\\

We still need to find the value of constant $A$. This can be done
by using the experimental fact about a smallest amount of energy
available at the microscale (condition $2$ of this section). To
this end we consider the relativistic Hamilton-Jacobi equation for
a massless particle (in itself a rather strange, but still valid,
concept within the framework of classical mechanics):
\begin{equation}
\label{S17} (\frac{\partial S}{\partial t})^2-(\nabla S)^2=0
\end{equation}
where we set the speed of light $c=1$.\\

One can easily see that it has two different solutions. One, let's
call it particle-like, is
\begin{equation}
\label{S18} S_p=-Et+\vec{p}\centerdot\vec{x}
\end{equation}
Another one, let's call it  wave-like, is
\begin{equation}
\label{S19} S_w=exp[-i(\omega t-\vec{k}\centerdot\vec{x})]
\end{equation}
On one hand, from (\ref{S18})
 \begin{equation}
 \label{S20} \frac{\partial
S_p}{\partial t}=-E
\end{equation}
 and from (\ref{S19})
\begin{equation}
\label{S21}
 \frac{\partial}{\partial t}(\frac{1}{i}Ln S_w)=-\omega
\end{equation}\

On the other hand, according to Planck's hypothesis about a
discrete character of energy transfer, we replace in (\ref{S20})
(for a single massless particle) energy
 $E$ by $\hbar\omega$, which yields
\begin{equation}
\label{S22} \frac{\partial }{\partial t}(\frac{S_p}{\hbar})=
-\omega
\end{equation}

From equations (\ref{S21}) and (\ref{S22}) immediately follows the
unique relation between two solutions, $S_p$ and $S_w$:
\begin{equation}
\label{S23} S_p=\frac{\hbar}{i}Ln S_w\equiv\frac{\hbar}{i}Ln\Psi
\end{equation}
As an additional bonus, by comparing $$\nabla S_p=\vec{p}$$ and
$$\nabla(\frac{1}{i}Ln S_w)=\vec{k}$$ we find from (\ref{S23}) the
De Broglie formula
\begin{equation}
\label{S24}
 \vec{p}=\hbar\vec{k}
 \end{equation}
 Thus the dual character ( wave-like and particle-like) of a
solution to the Hamilton-Jacobi equation inevitably leads to the
emergence of the $complex-valued$ wave
 "action" $S_w$ (wave function $\Psi$) related to the particle
action $S_p$ via a naturally arising substitution (\ref{S23}).\\

A comparison of (\ref{S13a}) and (\ref{S23}) allows us to
determine the value of constant $A$ in (\ref{S13a}):

\begin{equation}
\label{S25} A=\frac{\hbar}{i}
\end{equation}
which means that the relation between the auxiliary function
$S^{(q)}$ and the function reflecting both particle and wave-like
character of the phenomena on a microscale is
\begin{equation}
\label{S26} S=\frac{\hbar}{i}Ln\Psi
\end{equation}
The obtained relation provides $a ~priori$ the physical
justification of the substitution (\ref{S13a}) used by Schr$\mathrm{\ddot{o}}$dinger.\\

If we use constant $A$ from equation (\ref{S25}) in  (\ref{S15}),
we obtain the unique value of the "viscosity" $\nu^{(q)}$:
\begin{equation}
\label{S26a}
 \nu^{(q)}=\frac{\hbar}{2im}
\end{equation}

Now it becomes clear why we call $\nu^{(q)}$ a "viscosity":
$\hbar/m$ has a dimension of kinematic viscosity. Inserting
(\ref{S25}) in the linear equation (\ref{S16}) we arrive at the
Schr$\mathrm{\ddot{o}}$dinger equation:
\begin{equation}
\label{37} i\hbar\frac{\partial\Psi}{\partial
t}=-\frac{\hbar^2}{2m}\nabla^2\Psi+U\Psi
\end{equation}\

Here we have to make one more comment. As we have pointed earlier,
the dissipative term, heuristically introduced  into the
Hamilton-Jacobi equation, does not play any role at the classical
scales. One can consider it as small perturbations which become
significant only at the micro-scales. This proposition is
confirmed by the following reasoning. Smallness of the dissipative
term as compared with the rest of the terms in either
Hamilton-Jacobi equation (\ref{35a}) or the second law of Newton
[written as (\ref{32})] is determined by its comparison on a
dimensional basis with the dynamic term $p^2/mL$. The "viscous"
term is
$$\frac{\hbar p}{mL^2}\sim \lambda_{db}\frac{p^2}{mL^2}$$
(where $L$ is the characteristic length and $\lambda_{db}=\hbar/p$
is the De Broglie wavelength). The ratio of the latter and the
former
$$\lambda_{db}/L$$ becomes negligible, when we are dealing with
classical phenomena. This is fully consistent with treating a
classical path as a geometrical optics limit $\lambda\rightarrow
0$
of the wave propagation.\\

Interestingly enough, the introduction of the "dissipative" term
(in a form of small perturbations) into the classical equations of
motion (with a subsequent transition to a probabilistic
description) is compatible with $fractalization$ of the
deterministically defined classical path (one-dimensional curve)
which gradually degenerates into a quantum
fuzzy  "path", whose Hausdorff dimension is $2$ \cite{AG2,RF,LA}.\\

Now establishing the fruitfulness of our approach, we can apply it
to more complicated forms of the Hamilton-Jacobi equation. First,
we introduce the dissipative term into the Hamilton-Jacobi
equation for a charged particle in an electro-magnetic field
\begin{equation}
\label{b} \frac{\partial S}{\partial t}+\frac{1}{2m}(\nabla
S-e\vec{A})^2+e\phi=0
\end{equation}
where $\vec{A}$ and $\phi$ are the vector and scalar potentials
respectively.\\

When we follow the procedure outlined above, we must keep in mind,
that now instead of the definition of momentum $\vec{p}=\nabla S$
we have to use the generalized momentum $\vec{p}=\nabla
S-e\vec{A}$:
\begin{equation}
\label{38} \frac{\partial S}{\partial t}+\frac{1}{2m}(\nabla
S-e\vec{A})^2+e\phi=\nu^q\nabla\centerdot(\nabla S-e\vec{A})
\end{equation}
where the "viscosity" $\nu^{(q)}$ to be determined. Using
substitution (\ref{S26}) in (\ref{38}) and performing some
elementary vector operations we arrive at the following

\begin{flushleft}
\begin{eqnarray}
\label{39} \Psi\{-i\hbar\frac{\partial\Psi}{\partial
t}+\frac{1}{2m}(e^2A^2\Psi-2\frac{\hbar}{i}e\vec{A}\centerdot\nabla\Psi)
+e\phi\Psi-\nonumber\\
\nu^q(\frac{\hbar}{i}\nabla^2\Psi-e\Psi\nabla\centerdot\vec{A})\}+
(\nabla\Psi)^2(\nu^q\frac{\hbar}{i}-\frac{\hbar^2}{2m})=0
\end{eqnarray}
\end{flushleft}

By requiring this equation to be linear we get the following value
of constant $\nu^q$
$$\nu^q=i\frac{\hbar}{2m}$$
which is exactly the same (Eq.\ref{S26a}) as in the previous case
of the Schr$\mathrm{\ddot{o}}$dinger equation for an electrically
neutral particle.
 Inserting this value back in
(\ref{39}) we arrive at the respective
Schr$\mathrm{\ddot{o}}$dinger equation:
\begin{equation}
\label{40} i\hbar\frac{\partial\Psi}{\partial
t}=\frac{1}{2m}(\frac{\hbar}{i}\nabla-e\vec{A})^2\Psi+e\phi\Psi
\end{equation}\\
\subsection{Variational Principle for the Shr\"{o}dinger Equation as a
 Requirement of the Existence of the Lower Bound on Energy}
Here we would like to discuss the principle of least action as
applied to the Schr$\mathrm{\ddot{o}}$dinger equation. Generally
speaking, dissipation introduces irreversibility into a system,
and, quoting M.Planck  \cite{MP}, "irreversible processes are not
represented by the  principle of least action". Therefore it seems
paradoxical that the introduction of dissipation into a classical
mechanical system (in a form analogous to the one encountered in
classical fluid mechanics) would allow us to use the principle of
least
action.\\

However, in the first place, the latter will be applied not to the
classical action, but to the complex-valued wave function $\Psi$
replacing the former. Secondly, and this is a crucial point, the
"dissipation" which we are discussing is of a $special~ type$, a
code name for the information spread, reflected in a broadening
of the spatial information density \\

We argue here, that the principle of least action in this case
represents a requirement for the quantum system to have a lower
bound on its energy. Let us consider the difference between the
total energy and the potential and kinetic energies in  classical
mechanics, as expressed in terms of the classical action
\begin{equation}
\label{s1} \Delta \epsilon=-\frac{\partial S}{\partial t}-(\nabla
S)^2- U
\end{equation}
In classical mechanics this difference is identical zero,
($\Delta\epsilon =0$) indicating an arbitrary choice of the zero
energy. In quantum case, this is not so anymore, since one of the
salient features of a quantum system is boundedness from below of
its hamiltonian (that is energy), which implies the well-defined
choice of its zero (ground state) energy which is not necessarily
equals to zero.\\

To formally describe this feature we replace $S$ by
$(\hbar/i)Ln\Psi$ (according to \ref{S23}), use $(\nabla
S)\centerdot(\nabla S^*)$ instead of $(\nabla S)\centerdot(\nabla
S)$ to insure the real-valuedness of the respective term, and
define the difference $\Delta\epsilon$ as the following quantum
average:
\begin{flushleft}
\begin{eqnarray}\label{s2}
\Delta \epsilon =-\int\Psi^*[\frac{\partial S}{\partial
t}+(\nabla S)\centerdot(\nabla S^*)+ U]\Psi d^3q &=& \nonumber \\
 \int\Psi^*[\frac{\hbar}{i}\frac{1}{\Psi}\frac{\partial\Psi}{\partial
t}+\frac{\hbar^2/2m}{\Psi\Psi^*}(\nabla\Psi)\centerdot(\nabla\Psi^*)+U]\Psi
d^3q &=& \nonumber \\
-\int[\frac{\hbar}{i}\Psi^*\frac{\partial\Psi}{\partial
t}+\frac{\hbar^2}{2m}(\nabla\Psi)\centerdot(\nabla\Psi^*)
+U\Psi\Psi^*] d^3q &
\end{eqnarray}
\end{flushleft}

Now to satisfy the boundedness from below of the energy of a
quantum system we require the difference $\Delta\epsilon$
[represented by the functional (\ref{s2})] to have a $minimum$:
\begin{flushleft}
\begin{eqnarray}\label{s3}
\delta\int[\frac{\hbar}{i}\Psi^*\frac{\partial\Psi}{\partial
t}+\frac{\hbar^2}{2m}(\nabla\Psi)\centerdot(\nabla\Psi^*)
+U\Psi\Psi^*] d^3q & \equiv & \nonumber\\
-\int L(\Psi^*,\nabla\Psi^*,\Psi,\nabla\Psi,q,t)d^3q & = & 0
\end{eqnarray}
\end{flushleft}

 Here the Lagrangian $L$ is
\begin{equation}
\label{0} L=\frac{\hbar}{i}\Psi^*\frac{\partial\Psi}{\partial
t}+\frac{\hbar^2}{2m}(\nabla\Psi)\centerdot(\nabla\Psi^*)
+U\Psi\Psi^*
\end{equation}
Let us note that Lagrangian (\ref{0}) is usually introduced
heuristically like one of some possible choices (e.g.,\cite{LS}),
without referencing its physical meaning provided above.\\

Interestingly enough, the original solution of the problem of
quantization in micro-phenomena was treated by
Schr$\mathrm{\ddot{o}}$dinger \cite{S1} also as a variational
problem, albeit without indicating its physical meaning as the
requirement for the energy to have a minimum ( not necessarily
zero). In fact, Schr$\mathrm{\ddot{o}}$dinger wrote about his
awareness "that this formulation is not entirely unambiguous"
\cite{S1}. Our identification of the physical meaning of such a
variational principle  removes that ambiguity. Thus in terms of
$\Delta\epsilon$ we can represent the
Schr{$\mathrm{\ddot{o}}$}dinger's original variational problem  as
follows (taking into account that now $\Psi$ is a real-valued
function):
\begin{flushleft}
\begin{eqnarray}
\label{s6}
{\delta\int[(U-E)\Psi^2+\frac{\hbar^2}{2m}(\nabla\Psi)^2]d^3q\equiv}\nonumber\\
 -\delta\int\Psi[E-\frac{\hbar^2}{2m}\frac{1}{\Psi^2}
(\nabla\Psi)^2-U]\Psi d^3q=\nonumber\\-\delta<(\Delta\epsilon)>=0
\end{eqnarray}
\end{flushleft}

\subsubsection{Quantum Average of $\delta\epsilon$ for a
Quasi-Classical Limit of a Quantum Oscillator}

As an example of the variational problem  for the
Schr$\mathrm{\ddot{o}}$dinger equation as a requirement of the
lower bound on energy level we consider (\ref{s1}) for a
quasi-classical limit of the coherent state of the quantum
oscillator. In this case the wave function is
\begin{equation}
\label{s7} \bar{\Psi}= exp[\frac{1}{2}\alpha^2(\tilde{x}-cos\omega
t)^2-i(\frac{\omega t}{2}+\alpha^2\tilde{x} sin\omega
t-\frac{1}{4}b^2sin2\omega t)]
\end{equation}
where $a$ is the classical amplitude,~$\tilde{x}= x/a,
~\alpha=a\sqrt{m\omega/\hbar},~
\bar{\Psi}=\pi^{1/4}\alpha^{1/2}\Psi.$ Since now
$$S=\frac{\hbar}{i}Ln\Psi$$ and $$\tilde{x}\rightarrow cos\omega t$$we find
\begin{flushleft}
\begin{eqnarray}
\label{s8} \frac{\partial S}{\partial t}\rightarrow
-\frac{m\omega^2a^2}{2}-\frac{\hbar\omega}{2};\nonumber\\
\frac{1}{2m}\nabla S\centerdot\nabla S^*\rightarrow
\frac{m\omega^2}{2}a^2sin^2\omega t;\nonumber\\
U=\frac{m\omega^2}{2}a^2cos^2\omega t;\nonumber\\
-(\frac{\partial S}{\partial t}+\frac{1}{2m}\nabla
S\centerdot\nabla S^*+U)\rightarrow\frac{\hbar\omega}{2}
\end{eqnarray}
\end{flushleft}

This means that in the quasi-classical limit the quantum average
of this expression , that is "action"(which is actually not an
action, but  a difference between the total energy and the kinetic
and potential energies) given by the integral (\ref{s2}) is the
$ground~state$ ( read minimum) energy of the oscillator.
\subsubsection{Information Energy Density}

It is of interest to determine how much energy is required to
store(transmit) a unit of information in the case of a coherent
state. To this end we use the Lagrangian (\ref{0}) and find the
respective energy density $T_{00}$:
\begin{equation}
\label{sub1} T_{00}=\sum_k\frac{\partial\Psi_k}{\partial
t}\frac{\partial L}{\partial(\partial\Psi_k/\partial
t)}-L=-(\frac{\hbar^2}{2m}\nabla\Psi\centerdot\nabla\Psi^*+U\Psi\Psi^*)
\end{equation}
where $k=1,2$ and $\Psi_1=\Psi,\Psi_2=\Psi^*$. Upon substitution
the value of $\Psi$ from (\ref{s7}) in (\ref{sub1}) we obtain
\begin{equation}
\label{sub2} \frac{dE}{d\tilde{x}}=aT_{00}=
\frac{\alpha}{\sqrt{\pi}}\frac{m\omega^2a^2}{2}e^{-\alpha^2(\tilde{x}-cos\omega
t)}[2\tilde{x}(\tilde{x}-cos\omega t)+1]
\end{equation}

Dividing  (\ref{sub2}) by (\ref{s8}) we arrive at the expression
of information density with respect to the energy:
\begin{equation}
\label{sub3}
\frac{1}{\hbar\omega}\frac{dE}{dI}=\frac{2\tilde{x}(\tilde{x}-cos\omega
t)+1}{(1+Ln\sqrt{\pi})/\alpha^2+(\tilde{x}-cos\omega t)^2}
\end{equation}\
In the limiting case of the classical oscillator (\ref{sub3})
yields
\begin{equation}
\label{sub4} \lim_{\tilde{x}\rightarrow cos\omega
t}\frac{dE}{dI}=\frac{E_{cl}}{[(1+Log_2\sqrt{\pi})/2]}
\end{equation}
where $E_cl=m\omega^2a^2/2$ is the energy of the classical
oscillator. This indicates that approximately one bit of
information requires an expenditure of the classical energy of the
oscillator.\

 In another limit of very large values of $\tilde{x}>>1,
~\alpha\tilde{x}>>1$ we obtain from (\ref{sub3})
\begin{equation}
\label{sub5} \lim_{\tilde{x}\rightarrow\infty}\frac{dE}{dI}=
\hbar\omega
\end{equation}
which is exactly one bit of information per quantum of energy.\\

This  result is in full agreement with the conjecture
(\cite{BS},\cite{JB}) about a connection between an amount of
information $H$ transmitted by a quantum channel in a time period
$\epsilon\sim 1/\omega$ and energy  $E$ necessary for a physical
representation of the information in a quantum system
$$\frac{E\epsilon}{\hbar}\sim\frac{E}{\hbar\omega}\ge H$$
By setting $H_{min}=1$ we obtain our result (\ref{sub5}).\\
\begin{figure}
 \begin{center}
\includegraphics[width=6cm, height=6cm]{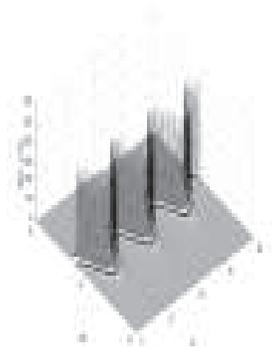}
\caption{\small  Energy density ( per unit of information) as a
function of $\tilde{x}=x/a$ and $t$ for $a/\lambda_{db} =20$}
 \end{center}
 \end{figure}\

\begin{figure}
 \begin{center}
\includegraphics[width=6cm, height=6cm]{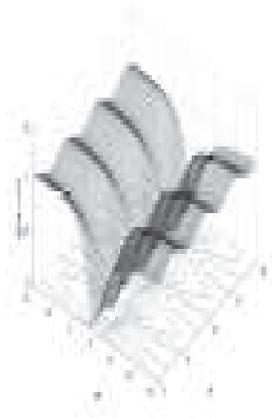}
\caption{\small  Energy density ( per unit of information) as a
function of $\tilde{x}=x/a$ and $t$ for $a/\lambda_{db} =0.5$}
 \end{center}
 \end{figure}\
 The graphs of the general distribution function $dE/dI$ for $2$
values of the parameter $\alpha=a/\lambda_{db}$:  $\alpha=20$
(classical regime) and $\alpha=0.5$(quantum regime) are shown in
Figures $7$ and $8$. It is seen that in the classical regime the
energy expenditure  per unit of information is very high in
classical regime is centered on the classical trajectory, while in
quantum regime this expenditure is "spread" over the space outside
the area occupied by the classical trajectory. This is in full
compliance with our previous discussion about the nature of
spatial spread of  information.

\subsection{Further Examples of Quantum Equations as a Consequence
of  Spatial Information Spread in Respective Classical Equations}

 As a next step, we apply the same idea  to a derivation of
the Klein-Gordon equation for a charged relativistic particle of
spin $0$ in an electro-magnetic field. To this end we add a small
perturbation term (analogous to the above "dissipative" terms
\footnote{the introduction of the full-blown dissipative term (as
in \cite{LL}) would lead to the emergence of a strongly nonlinear
equation, which still admits the solution proportional to
$exp(i\omega t$), and which we plan to address in the future}
$$\nu^q g^{jk}\frac{\partial}{\partial x^k}(\frac{\partial
S}{\partial x^j}+eA_j)$$ ($\nu^q$ is to be determined) to the
right hand side of the relativistic Hamilton-Jacobi equation
\begin{equation}
\label{23} g^{jk}(\frac{\partial S}{\partial
x^j}+A_j)(\frac{1}{m}\frac{\partial S}{\partial
x^k}+\frac{1}{m}eA_k)=m,
\end{equation}
(where we set the speed of light $c=1$) use  substitution
(\ref{S26}) and get
\begin{flushleft}
\begin{eqnarray}
\label{41} g^{jk}(\frac{\hbar}{i}\frac{\partial\Psi}{\partial
x^j}+e\Psi A_j)(\frac{\hbar}{im}\frac{\partial\Psi}{\partial
x^k}+\frac{e}{m}\Psi A_k) &=& \nonumber\\
m\Psi^2+\nu^qg^{jk}(\frac{\hbar}{i}\frac{\partial^2\Psi}{\partial
x^j\partial x^k}-\frac{\hbar}{i}\frac{\partial\Psi}{\partial
x^j}\frac{\partial\Psi}{\partial x^k}+e\Psi\frac{\partial
A_j}{\partial x^k})
\end{eqnarray}
\end{flushleft}

Linearity requirement imposed on this equation determines the
value of constant $\nu^q$: $$\nu^q=i\frac{\hbar}{m}$$ which is the
same as we found  before in Eq.(\ref{S26a}). Inserting this value
back in (\ref{41}) and performing some elementary calculations we
arrive at the Klein-Gordon equation for a charged relativistic
particle of spin $0$ in an electro-magnetic field:
\begin{equation}
\label{42}
 g^{jk}(\frac{\hbar}{i}\frac{\partial}{\partial
x^j}+eA_j)(\frac{\hbar}{i}\frac{\partial}{\partial
x^k}+eA_k)\Psi=m^2\Psi
\end{equation}\\

Since this idea clearly works for particles with zero spin, it is
naturally to ask whether it would work for particles with a spin.
Here one must be a little bit more ingenious in choosing the
appropriate dissipative term to be introduced into the
Hamilton-Jacobi equation. If we consider a classical  charged
particle in the electro-magnetic field it has an additional energy
$E_H=-\vec{\mu}\centerdot\vec{H}$ due to an interaction of the
magnetic
moment $\vec{\mu}$ and the magnetic field $\vec{H}$.\\

In terms of the vector potential $\vec{A}$ this energy is
\begin{equation}
\label{43} E_H=-\vec{\mu}\centerdot(\nabla\times e\vec{A})\equiv
div(\vec{\mu}\times e\vec{A})
\end{equation}

Experiments demonstrated that the magnetic moment $\vec{\mu_e}$ of
an electron is proportional to its spin $\vec{s}$:
\begin{equation}
\label{44} \vec{\mu_e}=\frac{\hbar}{m}\vec{s}
\end{equation}
It is remarkable that once again ( as in the above cases) the
coefficient of proportionality in (\ref{44}) has the dimension of
kinematic viscosity! Its magnitude is twice the magnitude of the
$"quasi~kinematic~viscosity"~ \nu^q$.\\

If we substitute (\ref{44}), in (\ref{43}) we obtain
\begin{equation}
\label{45} E_H=-\frac{\hbar}{m}\nabla\centerdot(\vec{s}\times
e\vec{A})
\end{equation}
This expression has a structure of the dissipative term introduced
earlier in the Hamilton-Jacobi equation (\ref{38}). Therefore we
rewrite this equation with the additional "dissipative" term
(\ref{45})
\begin{equation}
\label{46} \frac{\partial S}{\partial t}+\frac{1}{2m}(\nabla
S-e\vec{A})^2+e\phi=\nabla\centerdot[\nu^q(\nabla
S-e\vec{A})-\frac{\hbar}{m}(\vec{s}\times e\vec{A})]
\end{equation}\\

We substitute (\ref{S26}) in (\ref{46}), use the vector identity
$$\nabla\centerdot(\vec{a}\times\vec{b})\equiv\vec{b}\centerdot curl\vec{b}-\vec{a}\centerdot curl\vec{b}$$
 and obtain
\begin{eqnarray}\label{47}
-i\hbar\Psi\frac{\partial\Psi}{\partial
t}+\frac{1}{2m}(\frac{\hbar}{i}\nabla\Psi-e\vec{A}\Psi)^2+e\phi\Psi^2=\nonumber\\
\Psi^2\frac{\hbar}{i}\nu^q\nabla\centerdot(\frac{\nabla\Psi}{\Psi})-
\Psi^2e\frac{\hbar}{m}(\vec{A}\centerdot
curl\vec{s}-\vec{s}\centerdot curl \vec{A})
\end{eqnarray}
Since the required equation must be linear (which uniquely defines
$\nu^q$ again as $i\hbar/2m$), and  function $\Psi$ now depends on
the $z$-component of the spin $\vec{s}$ (that is, it becomes a
$2\times 1$ vector-column function) we have to replace vector
$\vec{s}$ by the respective ($2\times 2$) matrices
$\hat{\vec{s}}$. As a result, we arrive at the Pauli equation:
\begin{equation}
\label{48} \!i\hbar\frac{\partial\Psi}{\partial
t}=\frac{1}{2m}(\frac{\hbar}{i}\nabla-e\vec{A})^2\Psi+e\phi\Psi-
\frac{e\hbar}{m}(\hat{\vec{s}}\centerdot\vec{H})\Psi
\end{equation}\\

Since the  method of information spread introduced into the
classical Hamilton-Jacobi equations via the "effective viscosity"
$\nu^q$ has turned out to be fruitful so far, we apply it to a
simple case of a particle in the gravitational field. The
Hamilton-Jacobi equation in this case is
\begin{equation}
\label{49} g^{jk}S_{;j}S_{;k}-m^2=0
\end{equation}
where $g^{jk}$ is the metric tensor, $j,k=0,1,2,3$, the
$semicolon$ denotes covariant differentiation, and we set $c=1$.\\

Now we add to the right-hand side of (\ref{49}) the dissipative
term in the form used in the above calculations, that is
$\nabla\centerdot(\nu^q\nabla S)$. However, this time, instead of
the conventional derivatives, we use the covariant derivatives and
replace the constant scalar $\nu^q$ by a tensor function
$\nu^{jk}$. As a result, equation (\ref{49}) becomes:
\begin{equation}
\label{50} g^{jk}S_{;j}S_{;k}-m^2=(\nu^{jk}S_{;k})_{;j}
\end{equation}\\

By using  substitution (\ref{S26}) in (\ref{50}) and performing
some standard calculations we obtain the following
\begin{eqnarray}
\label{51} -\hbar^2g^{jk}\frac{\partial\Psi}{\partial
x^j}\frac{\partial\Psi}{\partial
x^k}-m^2\Psi^2+\frac{\hbar}{i}\nu^{jk}\frac{\partial\Psi}{\partial
x^j}\frac{\partial\Psi}{\partial
x^k}-\frac{\hbar}{i}\nu^{jk}_{;j}\Psi\frac{\partial\Psi}{\partial
x^k}\nonumber\\
-\frac{\hbar}{i}\nu^{jk}\Psi(\frac{\partial^2\Psi}{\partial
x^j\partial x^k}+\Gamma_{kj}^n\frac{\partial\Psi}{\partial
x^n})-m^2\Psi^2=0
\end{eqnarray}
where $\Gamma_{jk}^n$ is the Ricci tensor. We require this
equation to be linear, which uniquely determines the value of the
tensor $\nu^{jk}$:
\begin{equation}
\label{52} \nu^{jk}=i\hbar g^{jk}
\end{equation}
Since $g^{jk}_{;j}\equiv 0$ equation (\ref{51}) yields
\begin{equation}
\label{53} g^{jk}\frac{\partial^2\Psi}{\partial x^j\partial
x^k}-\frac{1}{\sqrt{-g}}\frac{\partial}{\partial
x^l}(\sqrt{-g}g^{nl})\frac{\partial\Psi}{\partial
x^n}+\kappa^2\Psi=0
\end{equation}
where $\kappa=m/\hbar$.\\

As a particular example we consider the centrally symmetric
gravitational field with the Schwarzchild metric:
\begin{eqnarray}
\label{54}
g^{jk}=0,j\neq k;~~~g^{00}=\frac{1}{1-r_g/r};~~~g^{11}=-(1-\frac{r_g}{r});\nonumber\\
g^{22}=-\frac{1}{r^2};~~~g^{33}=-(1-\frac{r_g}{r});~~~g=|g^{jk}|
=-\frac{1}{r^4\sin^2\theta};~~~r_g=2mG
\end{eqnarray}\
Equation (\ref{53}) is then
\begin{eqnarray}
\label{55} \frac{1}{1-r_g/r}\frac{\partial^2\Psi}{\partial
t^2}-(1-\frac{r_g}{r})\frac{\partial^2\Psi}{\partial r^2}-
\frac{1}{r^2
sin^2\theta}\frac{\partial^2\Psi}{\partial\phi^2}-\nonumber\\
\frac{1}{r^2}\frac{\partial^2\Psi}{\partial\theta^2}-\frac{2}{r}(1-\frac{3}{2}\frac{r_g}{r})
\frac{\partial\Psi}{\partial
r}-\frac{1}{r^2}cot\theta\frac{\partial\Psi}{\partial\theta}+\kappa^2\Psi=0
\end{eqnarray}

\section{Conclusion}

Physical phenomena can only be described as either particle-like
or wave-like phenomena. Consequently, the critical question
arises: Does the complex-valued wave function $\Psi$ represent
reality, or is it only an intricate device to deal with something
we don't have a complete knowledge of? \\

Bohm \cite{B} proposed to remove such indeterminacy and thus to
answer the above question by introducing hidden variables into the
existing Shr$\mathrm{\ddot{o}}$dinger equation.  We treat this
problem {\bf absolutely differently}, first
\begin{itemize}
\item by starting from the classical Hamilton-Jacobi equation
\footnote{Since the wave function is intrinsically connected to
the classical action, it seems appropriate to recall M.Planck's
words on the importance of action $S$ in physics. In his letter to
E.Schr\"{o}dinger \cite {LWM} he wrote,"I have always been
convinced that its ($action$) significance in physics was still
far from exhausted"}(without any presumed $a~ priori$ knowledge of
the Shr\"{o}dinger equation)
 and arriving at the Shr\"{o}dinger  equation,\

and secondly

\item without using any hidden variables, since they are not necessary
in such an approach.\

\end{itemize} Instead, and this is the major idea of our
approach, we demonstrate that the above indeterminacy is due to a
spread of information \footnote{the use of information in this
context is not surprising, since $S/\hbar^r$, where $r$ is the
number of the degrees of freedom, is roughly speaking the number
of quantum states, whose average negative logarithm represents the
system's entropy} over the whole space, with a simultaneous
preservation of information, in contradistinction to the classical
case where the same information is centered on
the classical path occupied by a classical particle.\\

Such an $information ~spread$  is described by an information
density function which is different from a delta-like function
observed in the classical case. In a sense, this resembles a
dissipation of temperature in a solid body, but only in a sense,
since the total quantity of heat remains unchanged. No wonder that
the resulting Shr\"{o}dinger equation has a form of the diffusion
equation, albeit with the imaginary "time" (which reflects the
intrinsic presence of wave features in this phenomenon). More to
the point, the above process resembles the transfer of energy from
larger to smaller scales in turbulence ( e.g., see Ref.
\cite{LL}).\\

 As a result, the wave-like quantum mechanics turns out to
follow from the particle-like classical mechanics due to the
explicit introduction in the latter of a dissipative mechanism
responsible for the spread of information. Consequently, the
initial precise information about the classical trajectory of a
particle is "spread" (but not lost) over the whole space, which
for a simple case of a spinless particle \cite{AG2}, \cite{LA}
results in a  transformation of a classical trajectory into a
fractal path with Hausdorff dimension of $2$ .\\

The idea of quantum mechanics representing a spatial spread of
information, implemented (in a general way) in this paper, has not
only made it possible to elementary derive the basic
quantum-mechanical equations from the continuum equations of
classical mechanics, but also seems to be applicable to more
complex and intriguing problems, as for example, a relativistic
particle in the gravitational field.\\

\newpage
{\bf{Acknowledgments}}.\

Author would like to thank Prof.V.Granik and G.Chapline for the
illuminating discussions and constructive criticism.

\end{document}